\documentstyle[prl,aps,preprint,floats,amsmath]{revtex}

\draft
\tighten

\input BoxedEPS
\SetRokickiEPSFSpecial  
\HideDisplacementBoxes

\newcommand{\half}{\fract{1}{2}}

\newcommand{\fract}[2]{{\textstyle\frac{#1}{#2}}}

\begin{document}

\title{Quantum Energies of Interfaces}
 \author{N.~Graham,\footnote{
e-mail:~graham@physics.ucla.edu,~jaffe@mit.edu,\\ maqua@mitlns.mit.edu,~herbert.weigel@uni-tuebingen.de}$^{\rm a}$
R.L.~Jaffe,$^{\rm b}$
M.~Quandt,$^{\rm b}$ and
H.~Weigel\footnote{Heisenberg Fellow}$^{\rm c}$\\
{~}\\$^{\rm a}$Department
of Physics and Astronomy~~ University of California at Los Angeles, Los
Angeles, CA  90095 \\
$^{\rm b}$Center for Theoretical Physics~~ Laboratory for Nuclear Science and Department of Physics,
Massachusetts Institute of Technology\\
Cambridge, Massachusetts 02139 \\
$^{\rm c}$Institute for Theoretical Physics~~ T\"ubingen University~~
Auf der Morgenstelle 14, D--72076 T\"ubingen, Germany\\
{~} \\
{\rm MIT-CTP\#3088 \qquad UNITU-HEP-07/2001 \qquad hep-th/0103010}
}

\maketitle

\begin{abstract}
We present a method for computing the one-loop, renormalized quantum
energies of symmetrical interfaces of arbitrary dimension and
codimension using elementary scattering data.  Internal consistency
requires finite-energy sum rules relating phase shifts to bound state energies.
\end{abstract}


Recent work in particle theory has highlighted the importance of a
class of time independent extended objects that are symmetric in $m$
``non-trivial'' spatial dimensions and independent of the coordinates
in the $n$ remaining, ``trivial'' spatial dimensions.  Examples
include domain walls in lattice field theories \cite{dwf,dwf2}, branes 
in string theory and extradimensional gravity \cite{branes}, and vortices 
and other phenomena in statistical mechanics. Generically, we refer to these 
objects as ``interfaces.''  In a series of earlier works, we developed 
methods to evaluate the renormalized, one-loop quantum contribution to 
the energy of solitions ($n=0$) \cite{method,1d,dimreg}. In this Letter, 
we extend our method to interfaces. We identify the one-loop quantum energy 
or equivalently the functional determinant or partition function of such 
a background with elementary quantities in scattering theory.  Our methods 
yield simple, unambiguous results renormalized in conventional
schemes, well suited to numerical computation.  Eq.~(\ref{11}), for
example, gives the one loop quantum energy for $n=1$ and $m<4$ in
terms of the partial wave phase shifts, the first and second Born
approximations, the energies of bound states and the Feynman two-point
function.  Extensions to arbitrary $n$ and $m$ do not become more
complicated in fundamental ways.

In the course of our analysis we find that the renormalizability of
the underlying field theory requires certain identities to hold within
the scattering data in $m$ dimensions.  These take the form of
``finite energy sum rules'' that generalize Levinson's theorem:
They relate integrals over the phase shifts,
regulated at high momentum by subtracting one or more Born
approximations, to the energies of bound states. The required
sum rules were first obtained by Puff within scattering theory some
time ago~\cite{Puff75}.  In Ref.~\cite{GJQW} we analyze the sum rules
in detail.  We generalize them to cases where more Born approximations
are subtracted than are demanded to regulate the high momentum piece,
we treat the symmetric channel in one spatial dimension, which is
anomalous, and we discuss how the sum rules can be understood as
generalizations of Levinson's theorem.  A related family
of subtracted sum rules has been obtained by Buslaev and
Faddeev~\cite{buslaev}.  However, these sum rules mix various orders
of the Born approximations and are thus not of particular use for
computing quantum energies.

For a static, pointlike object in $m$ dimensions described by a
classical background $\phi$, we would compute the ``effective
energy,'' $E_m[\phi]$, which is the effective action per unit time. In the present case, the relevant quantity is ${\cal E}_{n,m}[\phi]
=E_{n,m}[\phi]/L^{n}$, the effective energy per unit volume of the
trivial dimensions.  ${\cal E}_{n,m}$ looks like an interface tension
when viewed from the outside and like an induced cosmological constant
intrinsically.  It can be expressed as an infinite sum over (one
particle irreducible) Feynman diagrams where the fluctuating field
runs in a loop with all possible insertions of $\phi$, or equivalently
as a sum/integral over the shifts in the zero-point energies of the
fluctuating field in the background $\phi$.  Both of these expressions
are formally infinite for cases of interest.  We regulate these
divergences using dimensional regularization separately in $n$ and
$m$, use the tools of quantum mechanics to connect these two pictures,
and then renormalize ${\cal E}_{n,m}$ unambiguously.  We will take the
dynamical field $\psi$ to be either a complex boson or Dirac fermion
of mass $\mu$, coupled to the classical background $\phi$ by
$g\psi^{*}\phi\psi$ or $g\bar\psi\phi\psi$.  In the case of a
self-coupled scalar we employ a source to stabilize the classical
background if it is not a solution to the classical equations of
motion \cite{1d}.

Other approaches to this problem exist for the case $m=1$. Ref.~\cite{Yaffe} uses properties of one-dimensional functional
determinants to integrate over the non-trivial dimensions first (the
opposite order to us).  Ref.~\cite{Bordag} uses zeta function
regularization and analytic properties of the scattering amplitudes to
rotate to the imaginary $k$ axis.  Both renormalize by subtracting a
local term to cancel divergences.  In contrast, we subtract the entire
Feynman graph, and then add it back in and renormalize using standard
techniques.  For the leading subtraction, the tadpole graph, there is
no difference, since it is entirely local.  But for further
subtractions, which become necessary in higher dimensions, local
subtractions involve an arbitrary scale, and thus are difficult to
relate to definite renormalization schemes (such as on-shell, or
$\overline{\rm MS}$).  Refs.~\cite{Yaffe} and \cite{Bordag} have not
been generalized beyond $m=1$, whereas the generalization is
straightforward in our case.  In the cases where a direct comparison
can be made ($m=1$ and renormalization of the tadpole graph only) all
three approaches yield superficially distinct expressions, but appear
to us actually to be equivalent.

The zero-point energies and phase shifts, which are central to our method,
are determined by solving the time independent Klein-Gordon or Dirac
equation in the background $V(x)=g\phi(x)$. This background is restricted 
such that the scattering data and in particular the associated Jost functions can be uniquely determined~\cite{GJQW}. The Born approximation is an expansion in $g$. So too is the expansion of the effective energy in terms of the Feynman diagrams, where each insertion carries a factor of the potential $V(\vec{q})=g\int d^{m}x e^{-i\vec q\cdot \vec x}\phi(\vec x)$.
By identifying orders in $g$ we rewrite the ultra--violet divergent contributions to the quantum energy as Feynman diagrams whose divergences are unambiguously canceled by counterterms~\cite{method,1d,dimreg,Schw_Baa}.

In the continuum the sum over zero-point energies becomes a sum over
bound state energies and an integral over $\omega(k,p) =
\sqrt{k^2+p^2+\mu^2}$, weighted by the density of states.  Here $k$ and $p$
refer to the magnitudes of the momenta in the non-trivial and trivial
dimensions, respectively.  The density of states factorizes into
$(L/2\pi)^n$ times the density of states $\rho_{m}(k)$ in the $m$
non-trivial dimensions.  Note that $\rho_{m}(k)$ is independent of $p$.  We
assume enough symmetry in the background $\phi$ that the scattering
problem in the non-trivial directions decomposes into a sum over partial
waves.  Then it is well known that the change in the density of states due
to $\phi$ can be written as $\pi\delta\rho_{m}(k)=\sum_\ell D^\ell_m
d\delta_m^\ell(k)/dk$, where $D^{\ell}_{m}$ is the degeneracy of the
$\ell^{\rm th}$ partial wave in $m$ dimensions.  By convention, we take
$\delta^{\ell}_{m}(k)$ to be the sum over both signs of the energy.

As in the pointlike case, $n=0$, it is necessary to soften the infrared
($k=0$ and $p=0$) behavior~\cite{dimreg}.  The first step is to use
Levinson's theorem in each partial wave,
\begin{equation}
        \int_{0}^{\infty} \frac{dk}{\pi} \frac{d}{dk}
        \delta^\ell_m(k) + \sum_j 1 = 0 \,.
        \label{sum0}
\end{equation}
We multiply it by $\half\mu(p)=\half\sqrt{p^{2}+\mu^{2}}$ and subtract it
from the formal expression for the Casimir energy.\footnote{For $m=1$, the
right-hand side of eq.~(\ref{sum0}) is modified to $1$, which will cancel
the contributions from the half-bound states at $\omega=\pm \mu$ that exist
in the free background in this case \cite{1d,Levinson}.}  Then we can write
the fundamental expression for the effective energy per unit volume,
%
\begin{align}
        {\cal E}_{n,m}[\phi] &=
        \pm \int \frac{d^n p}{(2\pi)^n} \sum_\ell D^\ell_m \biggl[
        \int_0^\infty \frac{dk}{2\pi}
        \left(\omega(k,p)-\mu(p)\right) \frac{d}{dk}
         \delta_m^\ell(k)\nonumber\\
        &\hspace*{2.5in}{}+\frac{1}{2}\sum_j
        \left(|\omega^\ell_{j,m} (p)|-\mu(p)\right) \biggr] + {\cal
        C}_{n,m}[\phi]
        \label{regularized}
\end{align}
%
where
$|\omega^\ell_{j,m}(p)|=\sqrt{-(\kappa^{\ell}_{j,m})^{2}+p^{2}+\mu^{2}}$
are the absolute values of the bound state energies and the
$\kappa^{\ell}_{j,m}$ are the absolute values of their (imaginary)
momenta.  The overall sign in eq.~(\ref{regularized}) is for bosons
and fermions respectively.  ${\cal C}_{n,m}$ represents the
contributions of Lagrangian counterterms necessary to cancel
infinities and enforce a particular renormalization
scheme.

The integrals in eq.~(\ref{regularized}) diverge in the cases of interest
and should be regularized throughout our analysis.  We employ dimensional
regularization, so we assume both $n$ and $m$ are chosen in regimes where
the integrals converge.  This is the case for $0<m+n<1$.  Subsequently we
analytically continue to physically interesting cases ($n,m$ integers).

Our procedure is to identify potentially divergent diagrams in the
effective action expansion with terms in the Born approximation to the
phase shift.  We subtract these terms under the $k$-integral in
eq.~(\ref{regularized}) and then add back in exactly what we subtracted,
this time as Feynman diagrams ${\cal F}_{n,m}[\phi]$, which we combine with
the counterterms in the standard way.  The renormalized Feynman diagram
contributions, $\overline{\cal F}_{n,m}[\phi] = {\cal F}_{n,m}[\phi] +
{\cal C}_{n,m}[\phi]$, are a straightforward piece of our result.  The
number of subtractions required will depend on how large we want to allow
the final space dimension $m+n$ to get.  For $m+n<3$, subtraction of the
tadpole graph is sufficient.  For $3<m+n<5$ the two-point function must be
subtracted, and so on.  (For fermions, we must include also contributions
from higher-order graphs that eventually simplify because the symmetries of
the interaction relate them to lower-order graphs.)

We begin this procedure by subtracting the first Born approximation
$\delta^{(1)\ell}_{m}(k)$ from the phase shift $\delta^{\ell}_{m}(k)$ and
adding back in the contribution of the tadpole graph, ${\cal
F}_{n,m}^{(1)}[\phi]$.  In Ref.~\cite{dimreg} we {\it proved\/} the two are
equal using dimensional regularization for $n=0$.  This proof extends in a
straightforward way to the case of $n>0$.  Thus the effective energy per
unit volume with one subtraction becomes
\begin{align}
	{\cal E}_{n,m}[\phi]
	&= \pm \int\!\!\!\frac{d^n p}{(2\pi)^n} \sum_\ell D_{m}^\ell\biggl[
	\int_0^\infty \!\!\!\frac{dk}{2\pi} (\omega(k,p) - \mu(p))
	\frac{d}{dk}  \left(\delta_{m}^\ell(k) - \delta_{m}^{(1)\ell}(k) 	\right)\nonumber\\
&\hspace{2.5in}{} + \frac{1}{2}\sum_j (|\omega^\ell_{j,m} (p)|- \mu(p))
\biggr]
	+ \overline{\cal F}_{n,m}^{(1)}[\phi]\, .
	\label{onesub}
\end{align}
The two terms in ${\cal E}_{n,m}$ given by eq.~(\ref{onesub}) should now be 
separately finite for $m+n<3$. Integrating over $p$, we obtain
\begin{align}
	{\cal E}_{n,m}[\phi]&=
	\mp\frac{\Gamma(-\frac{1+n}{2})}{2(4\pi)^{\frac{n+1}{2}}}
	\sum_\ell D_{m}^\ell \biggl[
	\int_0^\infty \frac{dk}{\pi} (\omega^{n+1}(k) - \mu^{n+1})
	\frac{d}{dk}\left(\delta_{m}^\ell(k) - \delta_{m}^{(1)\ell}(k)           \right)\nonumber\\
&\hspace{2.5in}{}+\sum_j (|\omega^\ell_{j,m}|^{n+1} - \mu^{n+1}) \biggr] +
	\overline{\cal F}_{n,m}^{(1)}[\phi]
	\label{res1}
\end{align}
which presents a puzzle: if we take $n\to 1$ (say with $m=1$), ${\cal
E}_{1,1}$ appears to diverge because of the pole in the gamma
function.  The divergence is spurious, so the quantity in brackets
must vanish for $n=1$.  Furthermore, since each partial wave is
independent, each must vanish separately. {\bf This is guaranteed by the sum rule~\cite{Puff75}}
\begin{equation}
\int_0^\infty \frac{dk}{\pi} k^2 \frac{d}{dk}\left(\delta_m^\ell(k) -
\delta_m^{(1)\ell}(k) \right) - \sum_j (\kappa^\ell_{j,m})^2  = 0\, .
	\label{sum1}
\end{equation}
With the aid of eq.~(\ref{sum1}) we can take the $n\to 1$ limit,
\begin{align}
	{\cal E}_{1,m}
	&=\pm\frac{1}{4\pi} \sum_\ell D^\ell_m \biggl[\int_0^\infty \frac{dk}{\pi}
	k\log\frac{\omega(k)^2}{\mu^2}
	\left(\delta_m^\ell(k) - \delta_m^{(1)\ell}(k)\right)\nonumber\\
&\hspace{2.5in}{}- \frac{1}{2} \sum_j
	(\omega^\ell_{j,m})^2\log\frac{(\omega^\ell_{j,m})^2}{\mu^2} +
	(\kappa^\ell_{j,m})^2 \biggr]\, .
	\label{e1}
\end{align}
Here we have adopted the renormalization condition that the counterterm
exactly cancels the tadpole graph~\cite{dimreg}. Finally note that the
arbitrary scale of the logarithms cancels because of eqs.~(\ref{sum0})
and~(\ref{sum1}).

To extend to higher dimensions, we need to make a second Born
subtraction and add back the Feynman two-point function, which will
suffice for dimensions with $m+n < 5$.  We can continue this procedure
indefinitely -- subtracting higher Born approximations and adding back
the appropriate Feynman diagrams, which are renormalized by local
counterterms.  Questions of renormalizability enter only if we include
{\it ab initio\/} the vertices associated with these new counterterms.

To avoid new infrared problems for $m=1$, we perform an additional
subtraction\footnote{See ref.~\cite{GJQW} for a thorough discussion of the infrared anomalies that occur for $m=1$.}, by subtracting eq.~(\ref{sum1}) divided by $2\mu(p)$ from
eq.~(\ref{onesub}), so that $\omega(k,p)-\mu(p)$ is replaced by
$\omega(k,p)-\mu(p)-k^{2}/2\mu(p)$ under the $k$-integral.  Next we
subtract the second Born approximation, add back the Feynman two-point
function, ${\cal F}_{m,n}^{(2)}[\phi]$ and perform the $p$
integration:
\begin{align}
	{\cal E}_{n,m}[\phi]
	&=\mp\frac{\Gamma(-\frac{1+n}{2})}{2(4\pi)^{\frac{n+1}{2}}} \sum_\ell
	D^\ell_m\Bigg[\sum_j \Bigl(|\omega^\ell_{j,m}|^{n+1} - \mu^{n+1} +
	\frac{n+1}{2} (\kappa^\ell_{j,m})^2 \mu^{n-1} \Bigr)\nonumber\\
&\hspace{0.25in}{}+\int_0^\infty \frac{dk}{\pi} \Bigl(\omega(k)^{n+1} -
\mu^{n+1} -
	\frac{n+1}{2} k^2 \mu^{n-1} \Bigr)
	\frac{d}{dk} \left(\delta_m^\ell(k) - \delta_m^{(1)\ell}(k)
	- \delta_m^{(2)\ell}(k) \right)\Bigg]\nonumber\\
&\hspace{0.75in}{}	+ \overline{\cal F}_{n,m}^{(2)}[\phi]\, .
	\label{res2}
\end{align}
The coefficient of the gamma function now vanishes as $n\to 1$ by
construction.  The $n\to 1$ limit then gives
\begin{align}
	{\cal E}_{1,m}[\phi]&=
	\pm \frac{1}{4\pi} \sum_\ell\! D^\ell_m\!         \Bigg[\int_0^\infty\!\!\frac{dk}{\pi}
	k\log\frac{\omega(k)^2}{\mu^2} \left(
	\delta_m^\ell(k)-\delta_m^{(1)\ell}(k)-\delta_m^{(2)\ell}(k)\right)\nonumber\\
&\hspace{1.5in}{}-\frac{1}{2} \sum_j
	(\omega^\ell_{j,m})^2\log\frac{(\omega^\ell_{j,m}{})^2}{\mu^2} +
	(\kappa^\ell_{j,m})^2 \Bigg] + \overline{\cal F}_{1,m}^{(2)}[\phi].
	\label{11}
\end{align}
Eqs.~(\ref{e1}) and (\ref{11}) are identical for values of $m$ where
only one Born subtraction is necessary.  The contribution of the
second Born approximation has been replaced by the second Feynman
diagram.  However, eqs.~(\ref{res2}) and (\ref{11}) can be
continued to values of $n+m$ where two subtractions are necessary.

The finiteness of eq.~(\ref{res2}) as $n\to 3$ implies another scattering
theory identity,
\begin{equation}
	\int_0^\infty \frac{dk}{\pi} k^4
	\frac{d}{dk}\left(\delta_m^\ell(k) - \delta_m^{(1)\ell}(k) -
	\delta_m^{(2)\ell}(k) \right) + \sum_j (\kappa^\ell_{j,m})^4 = 0
	\label{sum2}
\end{equation}
which again is true channel by channel and can be derived directly
from scattering theory~\cite{Puff75}. In the limit $n\to 3$ we
then obtain
\begin{align}
	{\cal E}_{3,m}
	&= \pm \frac{1}{32\pi^2} \sum_\ell D^\ell_m
	\Bigg[-\int_0^\infty \frac{dk}{2\pi} 4 k \omega(k)^2 \log
	\frac{\omega(k)^2}{\mu^2} \Bigl(\delta_m^\ell(k)         -\delta_m^{(1)\ell}(k)-\delta_m^{(2)\ell}(k) \Bigr)\nonumber\\
&\hspace{0.75in}{} +
	\frac{1}{2} \sum_j
	\Bigl((\omega^\ell_{j,m})^4\log\frac{(\omega^\ell_{j,m})^2}{\mu^2} +
	\mu^2 (\kappa^\ell_{j,m})^2 - \frac{1}{2}(\kappa^\ell_{j,m})^4 \Bigr)
	\Bigg] + \overline{\cal F}_{3,m}^{(2)}[\phi] \, .
\end{align}

To illustrate our approach, we consider a dynamical scalar in a
background potential $V(x) = -\frac{\ell+1}{\ell} \mu^2 \mathop{\rm sech}^2
\frac{\mu x}{\ell}$.
\begin{figure}[hbt]
\centerline{\BoxedEPSF{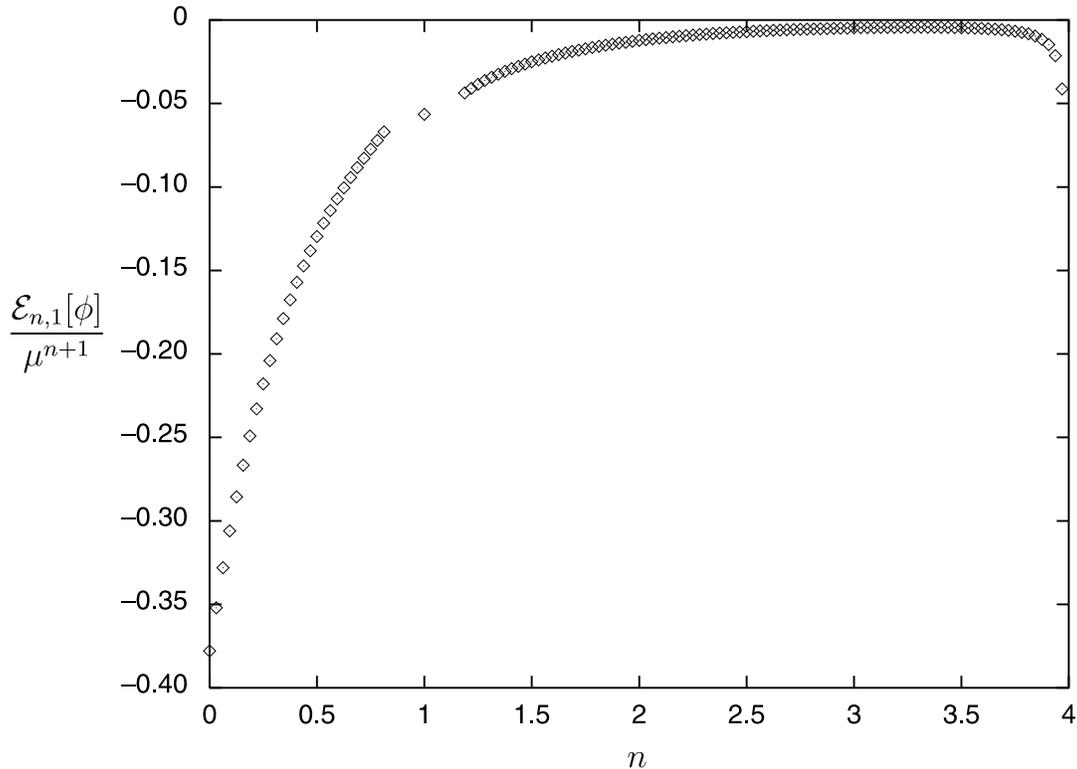 scaled 900}}\bigskip
\caption{\sl ${\cal E}_{n,m}[\phi]/\mu^{n+1}$ as a function of $n$ for a
bosonic field in the background $V(x) = -\frac{\ell+1}{\ell} \mu^2 \mathop{\rm sech}^2 \frac{\mu x}{\ell}$ with $m=1$, $\ell = 1.5$ and a
renormalization mass $M=\mu$.  For the particular cases $n=1$ and $n=3$,
the limits have been taken analytically using the sum rules in
eq.~(\ref{sum1}) and eq.~(\ref{sum2}).}
\label{nplot}
\end{figure}
For integer $\ell$, this potential is reflectionless and the phase
shifts can be obtained analytically, but we will take general $\ell$
and work numerically.  We perform two Born subtractions, and adopt the
renormalization conditions that the tadpole graph vanishes, and the
two-point function vanishes when the external four-momentum satisfies
$q^2 = M^2$.  For a background field $\phi$ of mass $M$, this choice
represents an on-shell renormalization scheme, which keeps fixed the
location of the pole in the two-point function.  (We do not perform
any wavefunction renormalization.)  Our approach allows us to hold
this scheme fixed as we vary either the background field or the number
of transverse dimensions $n$.  Figure \ref{nplot} shows the quantum
energy per unit volume in units of $\mu$ of the configuration for
$M=\mu$ and $\ell = 1.5$, as a function of $n$.  We have also used
these calculations to numerically verify the equivalence of the
expressions (\ref{e1}) and (\ref{11}).

We have described a simple, practical approach to computing quantum
energies of interfaces.  In the process, we have applied tools from
ordinary scattering theory that enabled us to precisely and
efficiently apply these techniques to a general class of physical
problems.

\emph{Acknowledgments} We thank U.~Wiese for suggesting a problem that 
led to these investigations.  We thank E.~Abers, E.~D'Hoker, J.~Goldstone, 
B.~M\"uller, and W.~Zakrzewski for helpful discussions, suggestions, and 
references. M.Q. and H.W. are supported in part by Deutsche
Forschungsgemeinschaft under contracts Qu 137/1-1 and We 1254/3-1. 
R.L.J. is supported in part by the U.S.~Department of Energy (D.O.E.) 
under cooperative research agreement \#DF-FC02-94ER40818. N.G. is supported 
by the U.S.~Department of Energy (D.O.E.) under cooperative research 
agreement \#DE-FG03-91ER40662.

\end{document}